\documentclass[twocolumn]{jpsj2} 

\def\runauthor{T. \textsc{Sasaki} {\it et al.}}

\title{%
Real space imaging of the metal - insulator phase separation in the band width controlled organic Mott system $\kappa$-(BEDT-TTF)$_{2}$Cu[N(CN)$_{2}$]Br
}

\author{%
T. \textsc{Sasaki}$^{1}$\thanks{E-mail address: takahiko@imr.tohoku.ac.jp}, N. \textsc{Yoneyama}$^{1}$, A. \textsc{Suzuki}$^{1}$, N. \textsc{Kobayashi}$^{1}$, Y. \textsc{Ikemoto}$^{2}$ and H. \textsc{Kimura}$^{2}$
}

\inst{%
$^{1}$Institute for Materials Research, Tohoku University, Katahira 2-1-1, Aoba-ku, Sendai 980-8577, Japan\\

$^{2}$SPring-8, Japan Synchrotron Radiation Research Institute, Mikazuki, Hyogo 679-5198, Japan\\
}
\recdate{\today}

\abst{%
Systematic investigation of the electronic phase separation on macroscopic scale is reported in the organic Mott system $\kappa$-(BEDT-TTF)$_{2}$Cu[N(CN)$_{2}$]Br.  Real space imaging of the phase separation is obtained by means of scanning micro-region infrared spectroscopy using the synchrotron radiation.  The phase separation appears near the Mott boundary and changes its metal-insulator fraction with the substitution ratio $x$ in $\kappa$-[($h$-BEDT-TTF)$_{1-x}$($d$-BEDT-TTF)$_{x}$]$_{2}$Cu[N(CN)$_{2}$]Br, of which band width is controlled by the substitution ratio $x$ between the hydrogenated BEDT-TTF molecule ($h$-BEDT-TTF) and the deuterated one ($d$-BEDT-TTF).  The phase separation phenomenon observed in this class of organics is considered on the basis of the strongly correlated electronic phase diagram with the first order Mott transition.  
}

\kword{%
Mott transition, phase separation, organic superconductor, $\kappa$-(BEDT-TTF)$_{2}$Cu[N(CN)$_{2}$]Br, infrared reflectance spectra, synchrotron radiation.
}

\begin{document}
\maketitle
\section{Introduction}

Microscopic spatially inhomogeneous electronic states have attracted much attention recently in many kinds of correlated electron systems.  
These electronic inhomogeneities have not been considered to be a heterogeneous mixture consisting of the multiphases with different chemical compositions such as the phase separation in the metal alloys.  
The materials with such intrinsic electronic inhomogeneity tend to possess a criticality of changes in charge, spin, orbital, and lattice degrees of freedom in the correlated electrons. 
Various types of the space conformation of the inhomogeneity and the response to the bulk properties have been found.  
Nano-scale spatial variation of the superconducting gap has been revealed in the superconducting state of Bi$_{2}$Sr$_{2}$CaCu$_{2}$O$_{8+\delta}$ by the scanning tunnelling spectroscopy and microscopy \cite{Lang}.  
In the normal state, charge carriers doped into antiferromagnetic insulators tend to group into some regions of the sample in the form of stripes in some copper oxides \cite{Tranquada}.
Meanwhile a different kind of the microscopic phase separation takes place in half-doped manganese oxides \cite{Fath}.  
Small variation from half doping causes phase segregation of electron-rich ferromagnetic and electron-poor antiferromagnetic domains with submicron size within the charge ordered phase.  
In the system with Mott transition, the nano-scale electronic inhomogeneity with preferred orientation has been found in slightly doped Mott insulator Ca$_{2-x}$Na$_{x}$CuO$_{2}$Cl$_{2}$ \cite{Hanaguri,Kohsaka}.

Organic charge transfer salts based on the donor molecule bis(ethylenedithio)tetrathiafulvalene, abbreviated BEDT-TTF or ET, have been recognized as one of the highly correlated electron systems.\cite{Michael} 
Among them, $\kappa$-(BEDT-TTF)$_{2}$$X$ with $X =$ Cu(NCS)$_{2}$, Cu[N(CN)$_{2}$]$Y$ ($Y =$ Br and Cl), etc. have attracted considerable attention from the point of view of the strongly correlated quasi two dimensional electron system because the native quarter filled band is modified to the effective half filled band by the strong dimer structure consisting of two BEDT-TTF molecules \cite{Kanoda,Kino,McKenzie1}.  
In such a strongly correlated electronic system, several electronic phases appear and the transitions among these phases are controlled by the applied pressure \cite{Lefebvre,Ito} and slight chemical substitution of the donor \cite{Kawamoto} and anion molecules \cite{Sushko}, which must change the conduction band width $W$ with respect to the effective Coulomb repulsion $U_{\rm dimer}$ between two electrons on a dimer \cite{Kanoda,Kino,McKenzie2}.  
The first order metal - insulator Mott transition divides the pressure - temperature plane of the phase diagram into the superconducting and antiferromagnetic (AF) Mott insulator phases.
The Mott transition line is extended to higher temperature beyond the superconducting transition temperature $T_{\rm c}$ and AF long-range ordering temperature $T_{\rm N}$ and terminated at the critical end point $T_{\rm cr} \simeq$ 40 K \cite{Lefebvre,Fournier,Limelette,Kagawa}.  
The phase diagram will be shown in Fig. 8.  
From $T_{\rm cr}$ to both weak and strong correlation sides in the phase diagram, the $T^{*}$ line and the bad metal - insulator line $T_{\rm ins}$ are elongated \cite{Sasaki1,Sasaki2,Mueller}.
In temperature above $T_{\rm cr}$, $T^{*}$, and $T_{\rm ins}$, the bad metallic state with the effective half filled band exists in wide range of the correlation strength which can be tuned by pressure and substitution of anion. 
In the weak correlation side from $T_{\rm cr}$, the bad metal changes to a correlated good metal through $T^{*}$ and then becomes superconducting.  
In the strong correlation side, the bad metal develops into a Mott insulator through $T_{\rm ins}$ and then becomes AF Mott insulator at $T_{\rm N}$.  
Thus the $\kappa$-(BEDT-TTF)$_{2}$$X$ family has been considered to be the band width controlled Mott system in comparison to the filling controlled one in the inorganic perovskites such as High-$T_{\rm c}$ copper oxides.

In addition to such the strong electronic correlation, characteristic flexibility in the molecules and the crystal lattices plays an important role cooperatively for modulating the electronic properties in $\kappa$-(BEDT-TTF)$_{2}$$X$. 
Particularly the terminal ethylene groups of BEDT-TTF molecules have conformational order-disorder glass transition at $T_{g} \simeq$ 80 K \cite{Mueller,Akutsu}.  
The degree of the disorder is expected to change with the cooling speed at $T_{g}$.  
Faster cooling may induce disorders in the terminal ethylene groups at low temperatures.  
This kind of disorders on molecular scale has been considered to have relation to the pinning effect of vortices \cite{Yoneyama1} and the quasiparticle scattering \cite{Yoneyama2}.  
Moreover such the cooling speed effect has been discussed on the change of the correlation strength $U_{\rm eff}/W$, which can switch the ground state of the sample located very near the Mott transition \cite{Taniguchi1}.

Inhomogeneous electronic states have been suggested in the $^{13}$C-NMR experiments near the first order metal-insulator transition in the artificially band width controlled $\kappa$-(BEDT-TTF)$_{2}$Cu[N(CN)$_{2}$]Br \cite{Miyagawa}.  
Below characteristic temperature $T^{*}$, $^{13}$C-NMR lines fall into two groups indicating the metallic and antiferromagnetic insulating nature.  
The results imply that two phases coexist spatially and statically.  
Subsequent transport experiments have suggested also such the coexistence of two phases at low temperature \cite{Taniguchi2}.  
Recently the real space imaging of the inhomogeneous electronic states has revealed the existence of the phase separation of the metal and insulator regions on micro-meter scale near the Mott transition of the band width controlled $\kappa$-(BEDT-TTF)$_{2}$Cu[N(CN)$_{2}$]Br \cite{Sasaki3}.  
Although it has been demonstrated that the phase separation phenomenon is realized near the Mott transition, the detail of the morphology, spatial distribution, size of domains and stability of the inhomogeneity have not been clarified yet.  
It is very important to obtain the detail of real space information which can give us a clue to know the mechanism of the macroscopic phase separation observed in the strongly correlated electronic system.  

In this paper we present the detail of the real space imaging of the electronic phase separation in the organic Mott system $\kappa$-[($h$-ET)$_{1-x}$($d$-ET)$_{x}$]$_{2}$Cu[N(CN)$_{2}$]Br, of which band width is controlled by the substitution ratio $x$ between the hydrogenated BEDT-TTF molecule ($h$-BEDT-TTF or $h$-ET) and the deuterated one ($d$-BEDT-TTF or $d$-ET). 
Scanning micro region infrared reflectance spectroscopy using the synchrotron radiation is applied to perform the two dimensional imaging of the local electronic state.  
This article develops in much greater detail of the recent our report\cite{Sasaki3} on this subject.  
Present paper contains the comprehensive results on the wide substitution $x$ dependence and the cooling condition dependence.  
On the basis of these results, the origin of the phase separation on macroscopic scale is discussed from the point of view of the electronic phase diagram.

\section{Experiments}

Single crystals of $\kappa$-[($h$-ET)$_{1-x}$($d$-ET)$_{x}$]$_{2}$Cu[N(CN)$_{2}$]Br partly substituted by deuterated BEDT-TTF molecule were grown by the standard electrochemical oxidation method \cite{Yoneyama3}.  
The typical dimensions of the samples obtained were $\sim 1 \times 1 \times 0.2$ mm$^{3}$.  
The largest rhombic surface corresponds to the conducting $c$-$a$ plane.
The substitution ratio $x$ denotes the nominal mole ratio to the fully deuterated BEDT-TTF molecule in the crystallization.
We checked the actual substitution with respect to the nominal value $x$ by measuring the intensity of the molecular vibrational mode of the terminal ethylene groups.
The substitution dependence of the macroscopic phase diagram and the superconducting properties have been examined \cite{Yoneyama3}.  
The full volume of the superconductivity has been observed in the range of $x = 0 - 0.5$ when the samples are cooled slowly.  
In this substitution range, $T_{\rm c}$ is increased slightly by about 0.1 K with $x$ up to $\sim$ 0.1.
Then $T_{\rm c}$ is kept at constant value of 11.9 K with $x$ up to $\sim$ 0.5.
Above $x = 0.5$, however, the superconducting volume fraction decreases and becomes about a few ten vol\% at $x = 1$, of which value strongly depends on the cooling condition.  
Concurrently with the reduction of the superconducting volume fraction, $T_{\rm c}$ starts to decrease from 11.9 K in $x =$ 0.5 to 11.6 K in $x =$ 1.  
The rest of the superconducting part may be the AF Mott insulator, and then magnetic impurity effect coming from such the coexistence may reduce $T_{\rm c}$. 
This simple substitution makes it possible to control the band width continuously at ambient pressure with minimal disorder effect. 
In rapid cooling, however, both the superconducting volume fraction and $T_{\rm c}$ decreases considerably with $x$.  
This superconducting volume fraction deduced from the magnetization measurements will be shown in Fig. 4 together with the results in the present paper.

Scanning micro region infrared reflectance spectroscopy (SMIS) measurements using the synchrotron radiation (SR) were performed at BL43IR in SPring-8\cite{Kimura,Ikemoto}.  
The polarized reflectance spectra were measured on the $c$-$a$ plane along $E \parallel a$-axis and $E \parallel c$-axis with a Fourier transform spectrometer (Bruker, 120HR/X) using KBr beam splitter and a polarizer in the mid-infrared (IR) range by use of a mercury-cadmium-telluride detector at 77 K.  
An IR microscope with the controlled precision $x$-$y$ stage and high brilliance of SR light enable us to obtain the two dimensional reflectance spectrum map with the spatial resolution of $\sim$ 10 $\mu$m with no apertures \cite{Kimura}.
The sample was fixed by the conductive carbon paste on the sample holder with a gold mirror which was placed at the cold head of the helium flow type refrigerator.  The window of the cryostat was BaF$_{2}$.
We gave careful consideration of less stress and good thermal contact to the crystals in the sample setting.  
In the normal cooling condition, the samples were cooled slowly from room temperature to 4 K in the rate of about 0.4 K/min. 
Other conditions in the rapid cool and thermal cycle experiments will be mentioned at the corresponding part in the text.
The reflectivity was obtained by comparison to that of the gold mirror at each temperature measured.
It took about one minute to acquire the IR spectrum at each scanning point in the case of the spectral resolution of 2 cm$^{-1}$ and accumulation of 100 times.  
Thus the total measurement time was about 8 hours for obtaining one imaging map ($\sim 0.3 \times 0.3$ mm$^{2}$).


\section{Results}

\begin{figure*}[tb]
\begin{center}
\includegraphics[viewport=0cm 9cm 20cm 26cm,clip,width=0.95\linewidth]{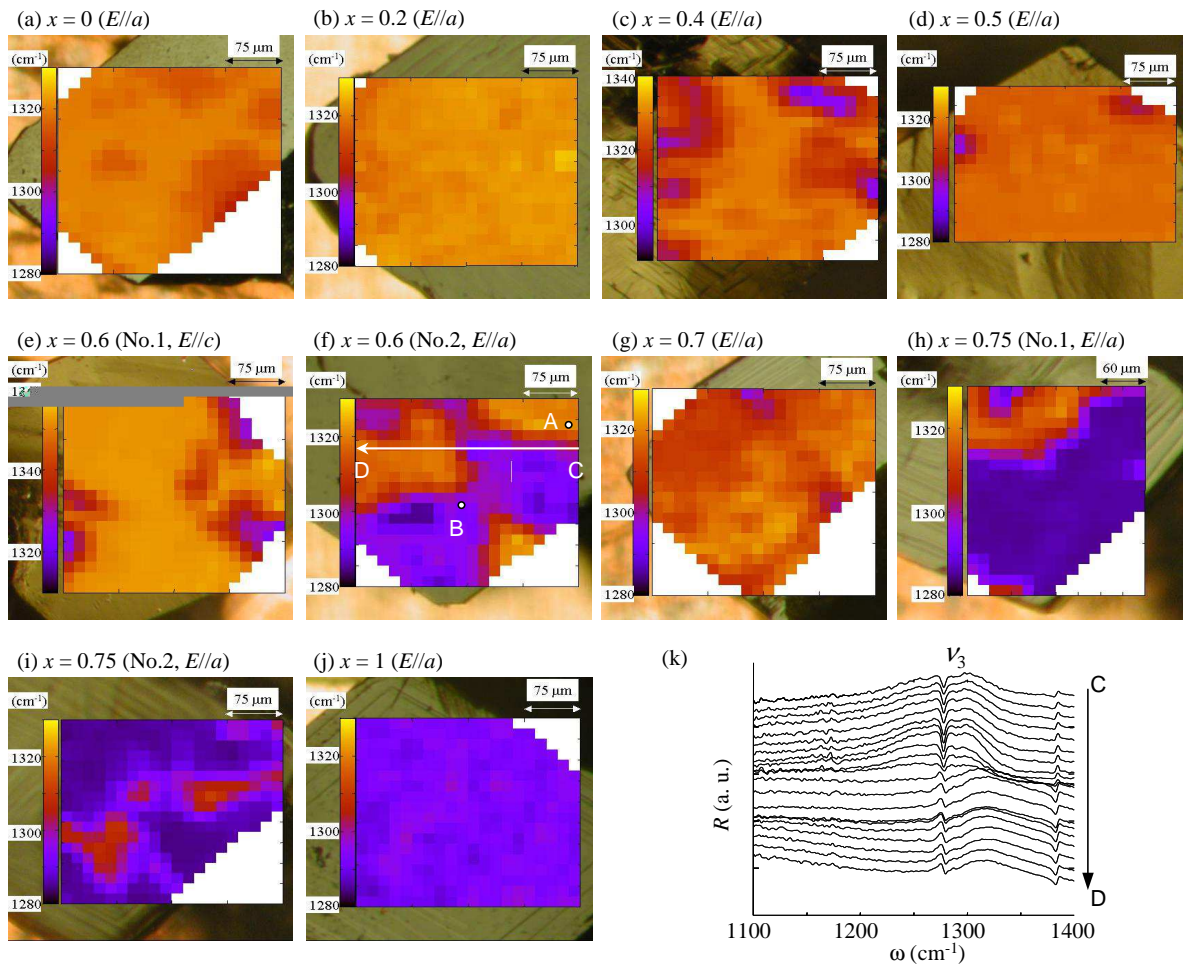}
\end{center}
\caption{Two dimensional contour map of the peak frequency of the $\nu_{3}(a_{g})$ mode on the crystal surface of $\kappa$-[($h$-ET)$_{1-x}$($d$-ET)$_{x}$]$_{2}$Cu[N(CN)$_{2}$]Br in (a) $x =$ 0 ($E \parallel a$), (b) $x =$ 0.2 ($E \parallel a$), (c) $x =$ 0.4 ($E \parallel a$), (d) $x =$ 0.5 ($E \parallel a$), (e) $x =$ 0.6 No.1 ($E \parallel c$), (f) $x =$ 0.6 No. 2 ($E \parallel a$), (g) $x =$ 0.7 ($E \parallel a$), (h) $x =$ 0.75 No.1 ($E \parallel a$), (i) $x =$0.75 ($E \parallel a$) and (j) $x =$ 1 ($E \parallel a$) at 4 K.  Bright color (higher frequency) indicates the metallic nature and dark color (lower frequency) does the insulating one. Spectra taken at the points A (metallic region) and B (insulating region) in Fig.1(f) are shown in Figs. 2(a) and 2(b), respectively.  (k) Reflectance spectra around 1300 cm$^{-1}$ along the arrow C-D in Fig. 1(f). Each spectrum is taken at 15 $\mu$m intervals. }
\label{fig1}
\end{figure*}

In order to obtain the real space image of the electronic states by SMIS measurements, we use the shift of the frequency $\omega_{3}$ of a molecular vibration mode $\nu_{3}(a_{g})$.  
The specific $\nu_{3}(a_{g})$ mode, which is a symmetric stretching mode of the central double bonded carbon atoms of BEDT-TTF molecule, has been found to be very sensitive to difference between metallic and insulating states in $\kappa$-(BEDT-TTF)$_{2}$$X$ \cite{Sasaki2} due to the large electron-molecular vibration coupling at the BEDT-TTF dimer\cite{Jacobsen,Rice}. 
The peak of the $\nu_{3}(a_{g})$ mode should shift to lower frequency in sharper shape in the insulating state at low temperature, while it shows opposite feature in the metallic state.  
We have obtained the real space imaging of the electronic phase separation in the $x =$ 0.5 and 0.8 samples of $\kappa$-[($h$-ET)$_{1-x}$($d$-ET)$_{x}$]$_{2}$Cu[N(CN)$_{2}$]Br by using such IR frequency shift of the $\nu_{3}(a_{g})$ mode \cite{Sasaki3}.

Figure 1 shows the two dimensional contour maps of the reflectivity peak frequency $\omega_{3}$ of $\nu_{3}(a_{g})$ mode at 4 K in (a) $x =$ 0 ($E \parallel a$), (b) $x =$ 0.2 ($E \parallel a$), (c) $x =$ 0.4 ($E \parallel a$), (d) $x =$ 0.5 ($E \parallel a$), (e) $x =$ 0.6 No.1 ($E \parallel c$), (f) $x =$ 0.6 No. 2 ($E \parallel a$), (g) $x =$ 0.7 ($E \parallel a$), (h) $x =$ 0.75 No.1 ($E \parallel a$), (i) $x =$ 0.75 No.2 ($E \parallel a$) and (j) $x =$ 1 ($E \parallel a$).  
An example of the shift of $\omega_{3}$ in space is shown in Fig. 1(k).  
Each reflectance spectrum is taken along the arrow C-D indicated in Fig. 1(f) at 15 $\mu$m intervals. 
The broad peak at about 1300 cm$^{-1}$ has been assigned to $\nu_{3}(a_{g})$ mode of $\kappa$-($h$-BEDT-TTF)$_{2}$Cu[N(CN)$_{2}$]Br \cite{Eldridge1}.  
Sharp dips around 1270 cm$^{-1}$ are the overlapping Fano type antiresonance of the terminal ethylene CH$_{2}$ vibration of the hydrogenated BEDT-TTF molecules.The $\nu_{3}$ mode shifts from lower to higher frequency around middle of the scanning path.  
The frequency shift occurs in relatively narrow region of about 1 - 2 scanning points corresponding to 15 - 30 $\mu$m.\cite{Sasaki3}
On the other hand, the CH$_{2}$ vibration mode does not change in the scanning, which is different from the $\nu_{3}$ mode.  
In the contour maps, bright region indicates the higher frequency of $\omega_{3}$ which demonstrates the metallic feature.\cite{Edep}  
Considering the experimental temperature of 4 K, the metallic region should be superconducting, which has been suggested by the bulk magnetization measurements \cite{Yoneyama3}.  
We cite, however, a metal instead of a superconductivity in the higher frequency region at 4 K because it is difficult to distinguish between two states in the present experimental method. 

The molecular vibration spectra of the $\nu_{3}(a_{g})$ mode observed at the metallic and insulating regions are compared in more detail with the spectra reported in the bulk superconducting and AF insulating samples.  
The spectra in Figs. 2(a) and 2(b) are obtained at the points A and B in the metallic and insulating regions of the $x =$ 0.6 No. 2 sample as shown in Fig. 1(f).
The peak shifts to lower frequency and becomes sharper at the point B, which indicates the insulating feature.  
Such spectra taken at the points A and B resemble those shown in Figs. 2(c) and 2(d), respectively, which have been obtained in the wider area ($\sim 0.5 \times 0.5$ mm$^{2}$) of the superconducting $\kappa$-($h$-BEDT-TTF)$_{2}$Cu[N(CN)$_{2}$]Br and AF Mott insulating $\kappa$-($h$-BEDT-TTF)$_{2}$Cu[N(CN)$_{2}$]Cl at 4 K by using the standard laboratory light source \cite{Sasaki2}.  
The magnitude of the antiresonance of the CH$_{2}$ vibration is reduced in the sample partly substituted by the deuterated BEDT-TTF molecules because the deuterated CD$_{2}$ vibration has a lower frequency of about 1040 cm$^{-1}$ \cite{Eldridge2}.  
Possible chemical inhomogeneity such as segregation of deuterated BEDT-TTF can be excluded by checking the magnitude of the antiresonance of the CH$_{2}$ vibrations at each scanning point. 

\begin{figure}[tb]
\begin{center}
\includegraphics[viewport=4cm 2cm 20cm 27cm,clip,width=0.6\linewidth]{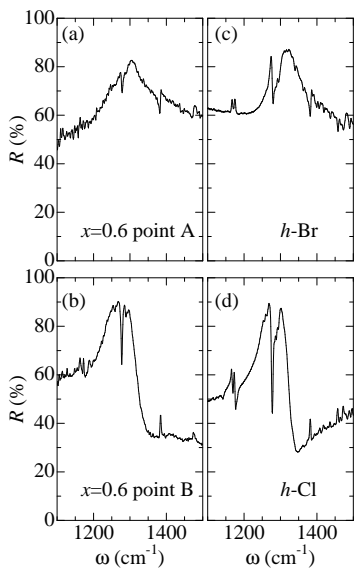}
\end{center}
\caption{Reflectance spectra ($E \parallel a$) around 1300 cm$^{-1}$.  (a) and (b) the spectra at the points A and B of $x=$ 0.6 No. 2 sample shown in Fig. 1(f), respectively. (c) and (d) the spectra in $\kappa$-($h$-BEDT-TTF)$_{2}$Cu[N(CN)$_{2}$]Br and $\kappa$-($h$-BEDT-TTF)$_{2}$Cu[N(CN)$_{2}$]Cl, which have been measured in wider area by using the laboratory light source \cite{Sasaki2}. }
\label{fig2}
\end{figure}

In the samples with small substitution ratio of $x =$ 0 and 0.2, almost homogeneous metallic state is realized in their whole area.  
Contrary to this, the sample of the opposite end member $x =$ 1 shows the homogeneous insulating state.  
In the intermediate substitution, however, inhomogeneous features on a micro-meter scale can be seen in the maps.  
The insulating domains appear on the dominant metallic background in the $x =$ 0.4 -- 0.5 sample.  
Such insulating domains have a tendency to grow with increasing $x$.  
In the $x =$ 0.75 sample, the insulating region becomes dominant and the metallic domains are found to be remained.  
The inhomogeneous electronic structure observed by the SMIS measurements has been concluded to be the electronic phase separation on the macroscopic scale.\cite{Sasaki3}  
 
The shape of the metallic and insulating domains is almost circle in case that those domains are small portion of the whole \cite{Sasaki3}.  
On one hand, the configuration of the metal and insulator regions, for example, in the intermediate $x =$ 0.6 - 0.75 sample is intricate and no specific size and orientation with respect to the crystal axes are observed.  
And each region seems not to be located at particular position such as the sample edge, step and scratch of the surface and so on. 
In addition, these structures and positions of each region are found to be stable to time at 4 K, which can be confirmed by getting almost the same image (Fig.7) in the $x =$ 0.75 No. 1 sample (Fig. 1 (h)) after 20 hours wait.  

\begin{figure}[tb]
\begin{center}
\includegraphics[viewport=3cm 2cm 18cm 29cm,clip,width=0.6\linewidth]{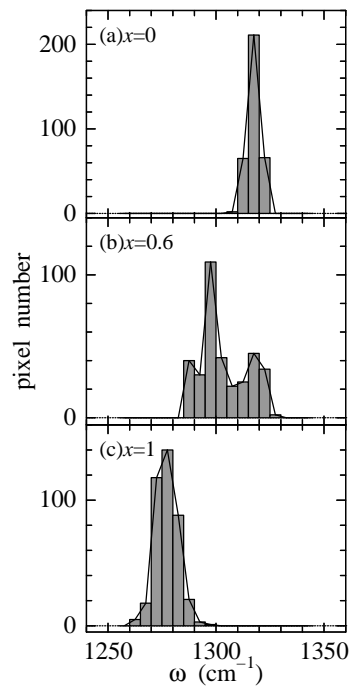}
\end{center}
\caption{Pixel number histogram for the $\nu_{3}(a_{g})$ mode frequency $\omega_{3}$ at 4 K in (a) $x=$ 0, (b) $x=$ 0.6 No. 2 and (c) $x=$ 1 samples.  }
\label{fig3}
\end{figure}

In order to estimate the metal - insulator volume fraction in the phase separation, the pixel number histograms for $\omega_{3}$ of (a) $x =$ 0, (b) $x =$ 0.6 No. 2 and (c) $x =$ 1 samples at intervals of 5 cm$^{-1}$ are shown in Fig. 3.  
In both the end members $x =$ 0 and 1 samples, the histograms show single symmetric narrow peak.  
In the $x =$ 0.6 No. 2 sample, however, $\omega_{3}$ distributes widely and indicates double peaks.
The ratio of the metallic and insulating regions is estimated from the subtotal pixel number of each lower and higher frequency peak.

\begin{figure}[tb]
\begin{center}
\includegraphics[viewport=3cm 13cm 18cm 24cm,clip,width=0.8\linewidth]{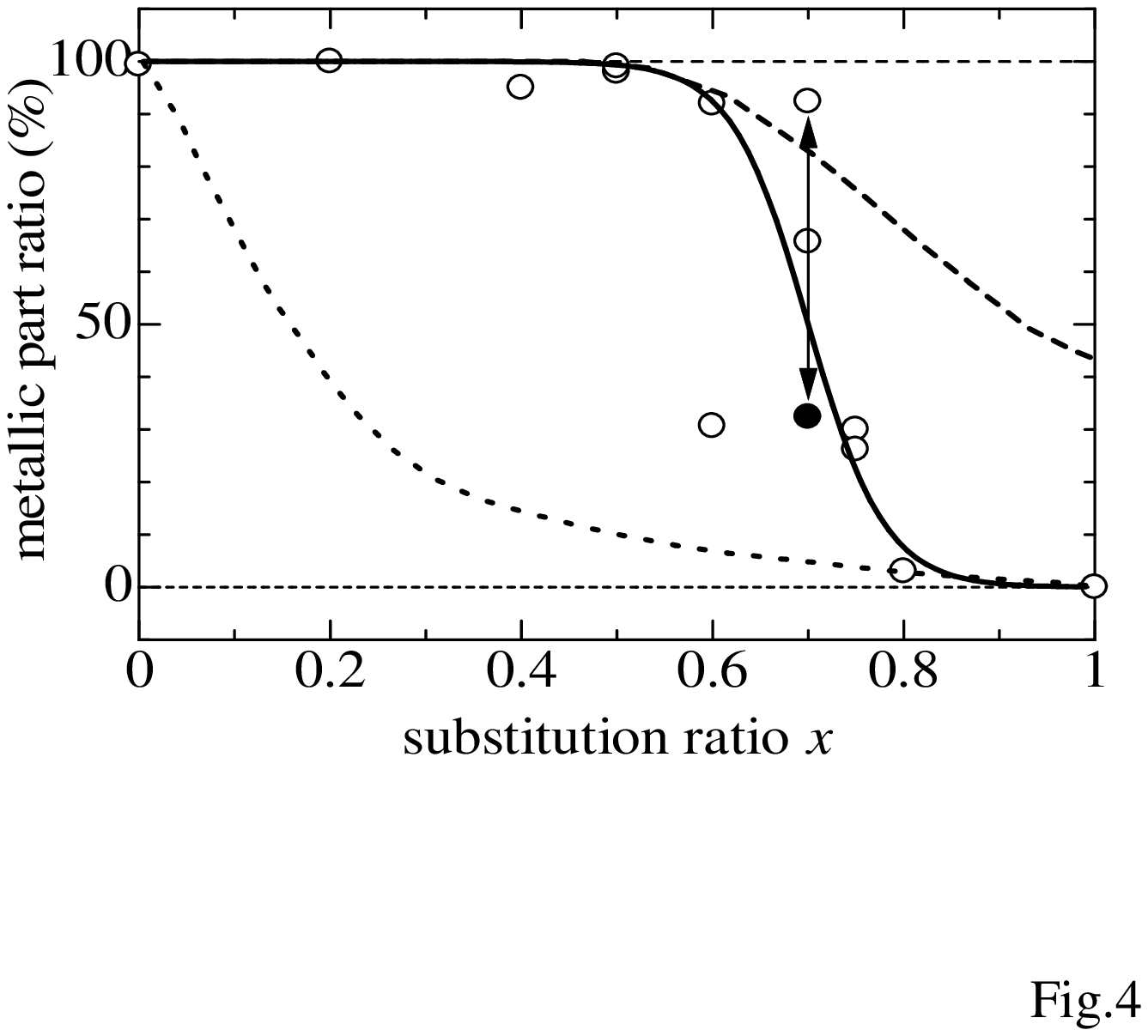}
\end{center}
\caption{Substitution ratio $x$ dependence of the metallic part ratio.  Open circles and filled one indicate the results of the slow cooled samples and the rapid cooled $x =$ 0.7 sample.  Solid curve is a guide for eyes.  Dashed and dotted curves represent the superconducting volume fraction deduced from the bulk magnetization measurements\cite{Yoneyama3} in the slow and rapid cooled conditions, respectively.}
\label{fig4}
\end{figure}

Figure 4 shows the substitution ratio $x$ dependence of the metallic part ratio deduced from the histogram for $\omega_{3}$ as shown in Fig. 3. 
Open circles indicate the results of the slow cooled samples. 
Previous results \cite{Sasaki3} and the data not shown in Figs. 1(a) -- 1(j) are included in Fig. 4.
A filled circle is the result obtained after the rapid cooled process in the $x = 0.7$ sample.  
The rapid cooled process will be explained latter in Fig. 5.
Solid curve is a guide for eyes.  
Dashed and dotted curves represent the superconducting volume fraction deduced from the bulk magnetization measurements \cite{Yoneyama3} in the slow and rapid cooled conditions, respectively.  
Almost full volume of the metallic nature is kept up to $x \simeq 0.5$ in both the present and magnetization experiments.  
The metallic part ratio starts to decrease rapidly above $x \simeq 0.5$ and indicates almost fully insulating at $x =$ 1 in the present SMIS results.
This decrease is rather steeper than the magnetization results.  
The difference may come from the following two reasons.  
One is the slight difference of the sample cooling conditions.  
The cooling speed in the present SMIS experiments is slightly faster than that in the magnetization experiments \cite{Yoneyama3}.  
Such different cooling conditions could modify the band width slightly in the same way as the deuterated BEDT-TTF substitution as is similar to the previous reports in $\kappa$-(BEDT-TTF)$_{2}$Cu[N(CN)$_{2}$]Br using the partly deuterated BEDT-TTF molecule.\cite{Taniguchi1,Taniguchi2}
Another is the difference of the detecting part of the samples in two experiments.  
The SMIS results should reflect a part of the surface with a thickness of several $\mu$m in depth.  
Although such micro-meter order thickness is enough for demonstrating the bulk properties, yet the results in the SMIS measurements may overestimate the insulating fraction near the surface. 

\begin{figure}[tb]
\begin{center}
\includegraphics[viewport=1cm 7cm 20cm 27cm,clip,width=0.9\linewidth]{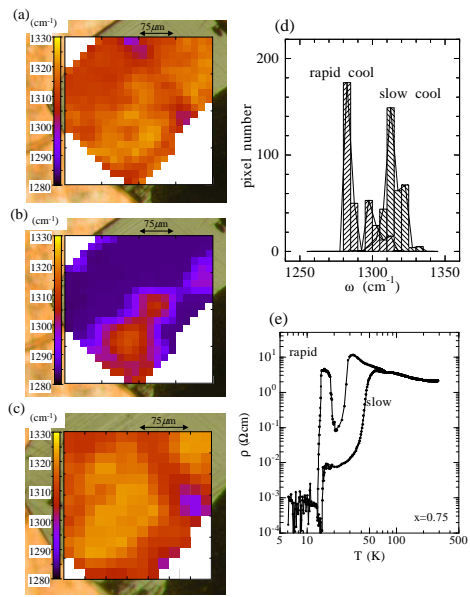}
\end{center}
\caption{Two dimensional contour maps of the peak frequency of the $\nu_{3}(a_{g})$ mode at 4 K of the same $x =$ 0.7 sample in Fig. 1(g).  The imaging were obtained in sequence as follows; (a) after slow cool at $\sim$ 0.4 K/min from room temperature to 4 K, (b) after rapid cool at $\sim$ 35 K/min from 120 K to 4 K, and (c) after slow cool at $\sim$ 1.4K/min from 100 K to 4 K.  (d) Frequency histogram of the $\nu_{3}(a_{g})$ mode at 4 K after slow and rapid cooled processes corresponding to Figs. 5(a) and 5(b), respectively. (e) The temperature dependence of the in-plane resistivity of a $x =$ 0.75 sample in slow and rapid cooled conditions. }
\label{fig5}
\end{figure}

\begin{figure}[tb]
\begin{center}
\includegraphics[viewport=3cm 2cm 16cm 27cm,clip,width=0.7\linewidth]{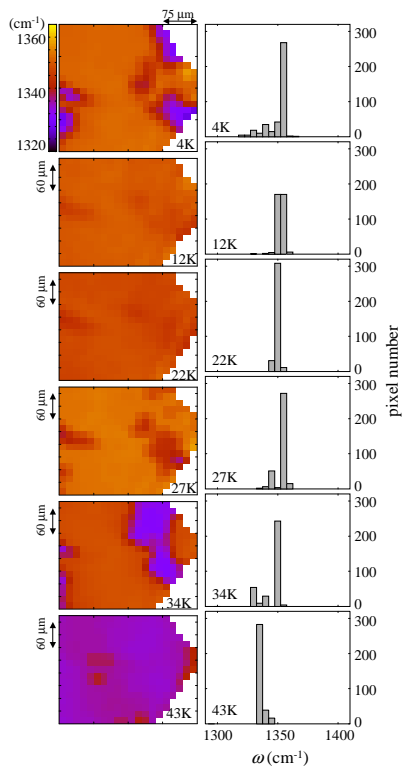}
\end{center}
\caption{Temperature variation of the two dimensional contour map of the peak frequency of the $\nu_{3}(a_{g})$ mode and the frequency histogram in the $x =$ 0.6 No. 1 sample. Imaging at 4 K is the same with Fig. 1(e). }
\label{fig6}
\end{figure}

\begin{figure}[tb]
\begin{center}
\includegraphics[viewport=3cm 1cm 16cm 27cm,clip,width=0.7\linewidth]{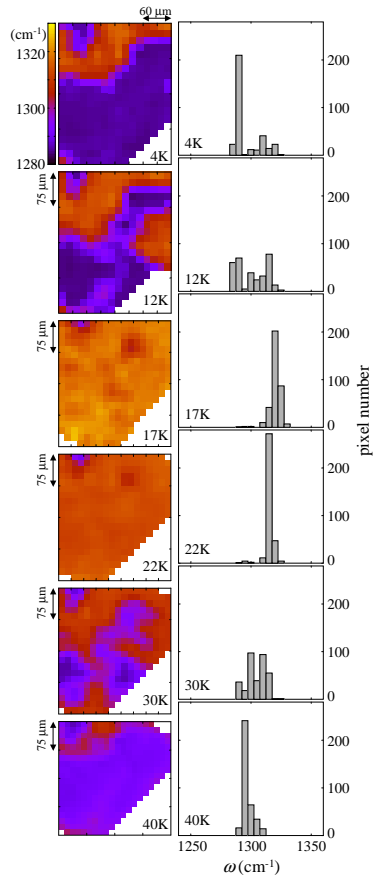}
\end{center}
\caption{Temperature variation of the two dimensional contour map of the peak frequency of the $\nu_{3}(a_{g})$ mode and the frequency histogram in the $x =$ 0.75 No. 1 sample. Imaging at 4 K is measured at the same area of Fig. 1(h) after 20 hours wait.  }
\label{fig7}
\end{figure}

As mentioned before, the metal-insulator fraction ratio in the phase separation is influenced by the cooling condition in addition to the substitution ratio $x$ in principle. 
Figures 5(a) - 5(c) show the variation of the phase separation with the different cooling conditions in the $x =$ 0.7 sample.  
Each imaging was performed in sequence after such the cooling processes as follows.
First the sample was cooled slowly ($\sim$ 0.4 K/min) from room temperature to 4 K (Fig. 5(a)). 
Second the sample was warmed up to 120 K and then rapidly cooled ($\sim$ 35 K/min) to 4 K (Fig. 5(b)).  
Finally the sample was warmed up to 100 K again, and then cooled slowly ($\sim$ 1.4 K) to 4 K (Fig. 5(c)).  
In the rapid cooled condition, the insulating fraction drastically increases as shown in Fig. 5(b).  
The metallic part ratio after such rapid cooled process is shown in Fig. 4 by the filled circle.  
The variation between the slow and rapid cooled results is indicated by the vertical arrow. 
The metallic fraction after the rapid cooled process is consistent with the magnetization experiments \cite{Yoneyama3}.  
The reduced metallic fraction is restored by getting warm up and cool the sample slowly again.  

The pixel number histogram of $\omega_{3}$ after slow and rapid cooled process in the $x =$ 0.7 sample is shown in Fig. 5(d).  
Shift to lower frequency side and the formation of the double peak structure of the histogram are found in the results of the rapid cooled state.  
These variation and reproducibility with the cooling speed and the thermal cycle suggest that the phase separation observed is not attributed to either the chemical inhomogeneity such as the segregation of the substituted $d$-BEDT-TTF molecules or the extrinsic causes such as the steps, scratches, dislocations on the crystal surface, but connected closely to the first order Mott transition.
The cooling condition dependence of the metal-insulator phase separation could be seen also in the resistivity behavior. \cite{Taniguchi1,Taniguchi2}
Figure 5(e) shows the temperature dependence of the in-plane resistivity of the $x = 0.75$ sample, which is a different sample used in SMIS measurements, in the slow (from 290 K to 5 K in 20 hours) and rapid (from 290 K to 5 K in 20 minutes) cooled conditions.  
Below $T_{g} \simeq$ 80 K, the resistivity in the rapid cooled condition shows an insulating behavior in contrast to the metallic behavior and superconductivity in the slow cooled condition.  
In the rapid cooled condition, reentrant insulator-metal-insulator transitions are seen at 30 and 18K, where the resistivity drastically changes by about two order of magnitude.  
The intrinsic transition width may be narrower than those observed experimentally because of the insufficient thermal equilibrium in the high cooling speed condition.  
The reentrant transitions represent a {\it S} shaped curve of the first order Mott transition line, which will be shown in Fig. 8.
A similar reentrance of the resistivity has been reported in $\kappa$-(BEDT-TTF)$_{2}$Cu[N(CN)$_{2}$]Cl under pressure.\cite{Ito}  
The superconducting transition at about 12 K is observed as the huge resistance drop and zero resistivity within the experimental resolution.  
The observed zero resistivity in the rapid cooled condition, however, should be attributed to a connection of percolating superconducting domains, which is demonstrated by the imaging of the phase separation in Figs. 5(a) - 5(c).  
In addition the drastic change of the resistivity at the insulator-metal-insulator reentrant transitions may reflect the phase separation near the Mott transition at higher temperature.

Figures 6 and 7 show the temperature variation of the two dimensional contour map of the peak frequency of the $\nu_{3}(a_{g})$ mode and the frequency histogram in the $x =$ 0.6 No. 1 ($E \parallel c$) and 0.75 No. 1 ($E \parallel a$) samples, respectively. 
Imaging at 4 K in the $x =$ 0.6 No. 1 sample is the same with Fig.1(e), but that at 4 K in the $x =$ 0.75 No.1 sample is measured at the same area of Fig. 1(h) after 20 hours wait.  
Both measurements have been performed from low to high temperature in sequence.  
The temperature variation of the phase separation in both $x =$ 0.6 and 0.75 samples are qualitatively the same as follows.
The metal-insulator phase separation observed at 4 K undergoes changes to a metal dominant state around 20 K and subsequently to a bad metal state above 40 K.  
Near the boundaries, which is identified latter with the $S$ shaped Mott transition curve, a formation process of the phase separation is observed, for example, at 27 and 34 K in the $x =$0.6 and at 12 and 30 K in the $x =$ 0.75 samples.  
The observation of such a reentrant process of the phase separation at both boundary of the $S$ shaped Mott transition curve again confirms the electron-correlation origin of the phase separation.

\begin{figure}[tb]
\begin{center}
\includegraphics[viewport=2cm 6cm 17cm 22cm,clip,width=0.9\linewidth]{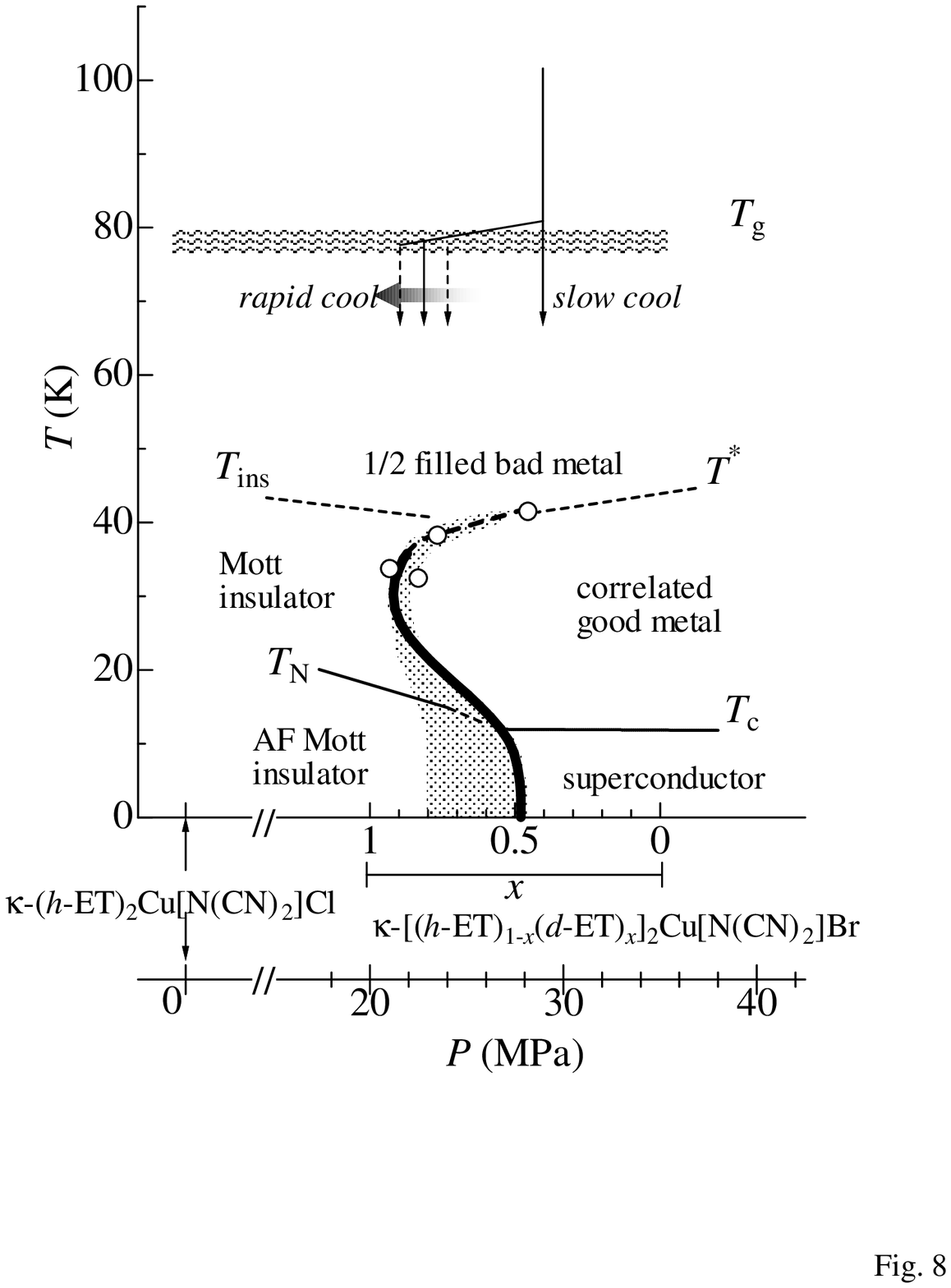}
\end{center}
\caption{Phase diagram of $\kappa$-(BEDT-TTF)$_{2}$$X$. Horizontal axis represents the substitution ratio $x$, and the corresponding pressure for $\kappa$-(BEDT-TTF)$_{2}$Cu[N(CN)$_{2}$]Cl in the slow cooled condition.\cite{Ito} Open circles indicate the critical end point of the first order Mott transition (thick solid line) in $\kappa$-(BEDT-TTF)$_{2}$Cu[N(CN)$_{2}$]Cl, which are extracted from literatures\cite{Lefebvre,Limelette,Fournier,Kagawa}. The phase separation appears in the shaded region. The horizontal position changes with cooling condition around $T_{\rm g} \simeq 75 - 80$ K.}
\label{fig8}
\end{figure}

\section{Discussion}

\subsection{Phase separation near the first order Mott transition}

We shall consider the relation between the phase separation observed and the phase diagram near the Mott transition.  
Figure 8 shows the schematic phase diagram of $\kappa$-(BEDT-TTF)$_{2}$$X$.  
The horizontal axis represents the substitution ratio $x$, and the corresponding hydrostatic pressure for $\kappa$-(BEDT-TTF)$_{2}$Cu[N(CN)$_{2}$]Cl in the slow cooled condition.\cite{Ito}
The deuterated BEDT-TTF substitution by ${\Delta}x =$ 0.1 corresponds to 1.5 MPa of the physical pressure.
The horizontal axis represents also the magnitude of the band width $W$ in comparison to $U_{\rm dimer}$.  
The band width increases in right side which corresponds to the weak electron correlation.  
Open circles indicate the critical end point of the first order Mott transition (thick solid line) in $\kappa$-(BEDT-TTF)$_{2}$Cu[N(CN)$_{2}$]Cl, which are extracted from literatures.\cite{Lefebvre,Limelette,Fournier,Kagawa} 
The critical end points are somewhat dependent on the experiments,\cite{Lefebvre,Limelette,Fournier,Kagawa} while the accompanying first order curves are the same with each experiment.  

The phase separation observed in the present experiments appears in the shaded narrow region of $x \simeq 0.5 - 0.8$ in $\kappa$-[($h$-ET)$_{1-x}$($d$-ET)$_{x}$]$_{2}$Cu[N(CN)$_{2}$]Br.  
In addition to the separation below the Mott transition in the insulator side at low temperature, it exists also near the transition at high temperature region.  
The observed reentrant behavior of the phase separation originates from the $S$ shaped curve of the Mott transition.
It should be emphasized that, considering from the Mott and superconducting transitions, the present partial substitution by the deuterated BEDT-TTF molecule can change the band width of the system continuously and homogeneously because these behaviors are in good agreement with the results in the partially deuterated BEDT-TTF molecule based $\kappa$-(BEDT-TTF)$_{2}$Cu[N(CN)$_{2}$]Br \cite{Kawamoto,Taniguchi1} and in $\kappa$-(BEDT-TTF)$_{2}$Cu[N(CN)$_{2}$]Cl under pressure.\cite{Lefebvre,Limelette,Fournier,Kagawa}  

The observed phase separation is considered to originate from the bistability of the free energy between a Mott insulator and a correlated good metal and the succeeding superconductor just on the first order Mott transition curve.  
When the system located in $x \simeq 0.5 - 0.8$ passes through the Mott transition curve at lower temperature part, the insulating domains arise in the correlated metal and superconducting matrix.  
The insulating ratio in the separation increases with increasing $x$.  
The process of the phase separation is considered not to be a spinodal decomposition but a nucleation and growth process.  
Because the pattern of the separation does not have characters of the spinodal decomposition, for example, the separated two regions tend to have a certain spacing and interconnected random pattern.\cite{Cahn}  

Coexistence and inhomogeneity of an AF - Mott insulating and a correlated metal-superconducting phase nearby the Mott transition have been suggested in pressurized $\kappa$-($h$-ET)$_{2}$Cu[N(CN)$_{2}$]Cl \cite{Lefebvre,Limelette,Kagawa}and $\kappa$-($d$-ET)$_{2}$Cu[N(CN)$_{2}$]Br at ambient pressure.\cite{Miyagawa}
Two different types of the coexistence region have been proposed although the observed coexistence/inhomogeneity may be the same with each experiment including present study.  
One is that the inhomogeneity appears below $T_{\rm cr}$ like as the present study,\cite{Limelette,Miyagawa} and another is that it does below $T_{\rm c}$ and $T_{\rm N}$.\cite{Lefebvre,Kagawa} 
This discrepancy may come from the different experimental methods to determine the phase boundary and the inhomogeneity.  
The transport measurements \cite{Limelette,Kagawa} should reflect the interconnection of the percolating metal-insulator domains, \cite{Uehara} which may be largely influenced by the history of the temperature and pressure variations.  
The ac susceptibility measurement \cite{Lefebvre} does not have enough sensitivity to distinguish between the Mott (paramagnetic) insulator and the good (paramagnetic) metal above $T_{\rm N}$ and $T_{\rm c}$.  
Therefore it can be concluded that the coexistence of the metal and insulator regions due to the phase separation accompanies the first order Mott transition and it appears not below $T_{\rm c}$ and $T_{\rm N}$ but below $T_{\rm cr}$.  

\subsection{Comparison of the inhomogeneity to other correlated electron systems}

The present electronic phase separation accompanying the phase transition may differ from the microscopic inhomogeneity reported in the transition metal oxides being irrespective of the macroscopic phase transition.  
Nanoscale inhomogeneity observed in the copper oxides appears in the carrier doping system to the Mott insulator, that is, in the filling controlled Mott system.  
In the pyrite type NiS$_{2-x}$Se$_{x}$ which shows a band width controlled metal-insulator transition by chemical substitution, however, the local electronic states are found to be spatially homogeneous even in the vicinity of the transition.\cite{Iwaya}  
In comparison to these results in the inorganic system, we consider that nano-scale inhomogeneity does not exist in the present organic band width controlled Mott system although a possibility of the existance of such a nano-scale structure inside each domains or each scanning spot is not excluded.
But close agreement between the previous results of NMR and magnetization and present results suggests that the phase separation occurs on macroscopic scale.
We therefore expect that the nano-scale intrinsic inhomogeneity appears in the filling controlled Mott system without thermodynamic phase transition, and the macroscopic inhomogeneity, if it comes, should accompany the critical phenomena like as the phase transition and non-equilibrium state.  
In a sense, the latter situation may be relevant to the phase separation/segregation in the antiferomagnetic insulator - ferromagnetic metal phase transition of the manganese oxides.\cite{Fath,Uehara}

\subsection{Superconductivity and the phase separation}

The superconductivity in $\kappa$-(BEDT-TTF)$_{2}$$X$ has been often discussed in comparison with that of the high-$T_{\rm c}$ copper oxides because of the similarity in appearance of the phase diagram and possible anisotropic $d$-wave pairing symmetry \cite{McKenzie1}.  
Then the most of the theoretical investigations have been based on the spin fluctuation mediated anisotropic superconductivity \cite{Kino2,Kondo,Kuroki,Schmalian,Louati,Kuroki2}.  
In this line the multicritical phenomenon of the superconductivity and AF insulator has been discussed on the basis of the SO(5) symmetry between two states \cite{Murakami,Onoda}.  
From the experimental points of view, however, the correlated good metal region and also the metallic domain in the phase separation do not give an indication of growing the AF fluctuation to superconductivity but restraining the fluctuation below $T^{*}$ to a good metal.\cite{Kanoda,Sasaki1,Miyagawa,Kawamoto2}  
This is different from the pseudogap state in the high-$T_{\rm c}$ copper oxides, which could be a precursor of the superconductivity.  
The domain structure consisting of the Mott insulator having the AF fluctuation and the correlated good metal in the phase separation below $T_{\rm cr}$ exists in the same way as the AF Mott insulating and superconducting domains below $T_{\rm N}$ and $T_{\rm c}$.  
This may demonstrate less communication between the Mott insulator and the good metal states and those do not interact with each other.  
Subsequently the similar independency between the superconductivity and AF Mott insulator may be retained.  

Recently a first order Mott transition from AF Mott insulator to a $d$-wave superconductor was found theoretically.\cite{Powell,Gan}
In one paper,\cite{Gan} however, no coexistence of the two phases was concluded within their calculation, while a possibility of a phase separation was proposed in the other approach.\cite{Capone}
We need further consideration of the superconductivity induced from the AF fluctuation in the Mott insulator region separated by the first order transition extended up to $T_{\rm cr} \gg T_{\rm N}$ and $T_{\rm c}$.  
It is noted that the other models for attractive interaction of pairing have been proposed.\cite{Girlando,Varelogiannis,Bill}
A model \cite{Varelogiannis} of the small-q phonon mediated superconductivity has reproduced also the anisotropic $d$-wave gap experimentally observed in $\kappa$-(BEDT-TTF)$_{2}$Cu(NCS)$_{2}$ \cite{Izawa,Arai} .

\subsection{Effect of the cooling speed at $75 - 80$ K on the phase separation}

It has been known that rapid cooling at $75 - 80$ K in $\kappa$-(BEDT-TTF)$_{2}$Cu[N(CN)$_{2}$]Br modifies the electronic state at low temperature due to the structural changes.
There are two types of the structural change at $75 - 80$ K.
One is the conformational order-disorder glass transition of the terminal ethylene groups in BEDT-TTF molecules.  
The ethylene groups are thermally activating above $T_{\rm g} \simeq 75 - 80$ K and should be frozen in the conformational ordered state below $T_{\rm g}$. \cite{Mueller,Akutsu} 
But some small number of ethylene groups remain as disorders with different conformation.  
These disorders may act as an impurity in the electronic state and the number of disorders is expected to increase in faster cooled condition.\cite{Yoneyama1,Yoneyama2,Stalcup}  
Another is the structural change at $75 - 80$ K, which is accompanied by the non-monotonic and anisotropic temperature dependence of the in-plane lattice parameters $a$ and $c$.\cite{YWatanabe}
This implies a change of the band structure at $75 - 80$ K.
Recent detailed structural study has found that the rapid cooled process decreases the inter-dimer transfer integral in comparison to the intra-dimer one, which leads to smaller $W/U_{\rm dimer}$, without increasing the disorders of the terminal ethylene groups.\cite{MWatanabe}
As far as the authors know, the relation between the disorder-order transition of the terminal ethylene groups at $T_{g}$ and the anomalous temperature dependence of the lattice parameters has not been understood yet.
Considering these results, however, the cooling condition dependence of the phase separation could be explained by the change of the correlation strength, that is, the horizontal position in Fig. 8 with the cooling speed around $T_{\rm g} \simeq 75 - 80$ K where the structural change takes place.
In this context, the disorders do not induce the insulating domains by themselves because amount of disorders must be a few \% of the BEDT-TTF molecules at most in slow cooing and $5 - 20$ \% even in the rapid cooled condition \cite{Mueller,Akutsu,MWatanabe}.  
Then the bistability of two phases near the Mott critical point may be essential for self-assembled metal-insulator domains and the disorders may have the side role as the nucleation, if any.  

\subsection{Macroscopic scale inhomogeneity in molecular crystals}

Similar inhomogeneity on micro-meter scale has been found in the current driven low resistive state of the quasi one dimensional organic Mott insulator K-TCNQ \cite{Kumai,Okimoto}.
In the low resistive state a visible stripe pattern composed of alternating dark and bright regions emerged with a spacing of 3 to 14 $\mu$m.  
The dark and bright regions could be viewed as the periodic phase segregation into carrier-rich (nondimerized) and -poor (dimerized) regions.
Although the pattern formation on micro-meter scale is realized in a non-equilibrium state, yet the current-induced coexistence of such the phase segregation has been concluded to be inherent to the correlated electron system where the hypothetical metallic state is adjacent to the Mott insulator.  

It is arguable that such metallic domains observed in the present and also the current driven phase separation is stable against the Coulomb interaction between metallic domains.  
We have no definite information to discuss this problem at present.  
It must be important to know the dielectric response in the Mott insulator state because the large enough dielectric constant could effectively screen the Coulomb interaction.  

A characteristic micro-meter scale in the electronic inhomogeneity has been also suggested by the nonlinear conduction in the insulating state, most probably charge ordered state, of $\theta$-(BEDT-TTF)$_{2}$CsZn(SCN)$_{4}$.\cite{Terasaki}  
From the analysis of the threshold electric field, the observed giant nonlinear conduction is assigned to a collective excitation with a coherent length scale of $\sim$ 1 $\mu$m.  
The authors \cite{Terasaki} expect that a percolated charge order with the length scale of $\sim$ 1 $\mu$m is distributed in the bulk sample inhomogeneously.

Molecular crystals can be considered to be a clean system, which is free from point defects, impurities, and so on in comparison to the inorganic compound like as oxides.  
The mean free path is long enough to observe the magnetic quantum oscillations at low temperature in most metals.\cite{Sasaki4,Kartsovnik}
Do inhomogeneous electrons near the critical boundary in such a clean system tend to have a conformation with a macroscopic micro-meter order length scale irrespective of the origins ?  
To clarify this question, space imaging technique with the resolution from nano- to micro-meter must be developed and applied to study in the critical and nonequilibrium states of the correlated clean electron system like as molecular crystals.

\section{Conclusion}

In conclusion, experimental evidence of the electronic phase separation is obtained by using the real space imaging technique on the single crystal surface of the organic Mott system $\kappa$-[($h$-ET)$_{1-x}$($d$-ET)$_{x}$]$_{2}$Cu[N(CN)$_{2}$]Br.  
SMIS measurements using SR enable us to show the macroscopic size of the domain structure of the insulating and metallic regions.  
This finding does not exclude the possibility of the nano-scale inhomogeneity inside each domains or each scanning spot because the obtained spectrum may result in the average of nano-scale inhomogeneity in the measured spot. 
But close agreement between the magnetization and present results suggests that the phase separation occurs on macroscopic scale.
The observation of the micro-meter scale phase separation is different from the recent findings of the nano-scale electronic inhomogeneity in the strongly correlated inorganic system.  
The origin of the phase separation may be the strong electronic correlation at the first order Mott transition.

\section*{Acknowledgment}

We are grateful to T. Moriwaki, T. Hirono and T. Kawase for their technical supports of SMIS measurements at SPring-8.
The authors thank M. Watanabe for useful discussions.
SR experiments were performed at SPring-8 with the approval of JASRI (2003B0114-NSb-np, 2004A0023-NSa-np and 2004B0014-NSa-np).
This work was partly supported by a Grant-in-Aid for Scientific Research (C) (Grant No. 15540329) from JSPS and Scientific Research on Priority Areas (Grant No. 1603824) from MEXT.


\begin{thebibliography}{99}

\bibitem{Lang} K.~M.~Lang, V.~Madhavan, J.~E.~Hoffman, E.~W.~Hudson, H.~Eisaki, S.~Uchida, and J.~C.~Davis: Nature (Jondon) {\textbf 415} (2002) 412.
\bibitem{Tranquada} J.~M.~Tranquada, B.~J.~Sternlieb, J.~D.~Axe, Y.~Nakamura, and S.~Uchida: Nature (London) {\textbf 375} (1995) 561. 
\bibitem{Fath} M.~F{\"a}th, S.~Freisem, A.~A.~Menovsky, Y.~Tomioka, J.~Aarts, and J.~A.~Mydosh, Science {\textbf 285} (1999) 1540.
\bibitem{Hanaguri} T.~Hanaguri, C. Luplen, Y.~Kohsaka, D.~-H.~Lee, M.~Azuma, M.~Takano, H.~Takagi, and J.~C.~Davis, Nature (London) {\textbf 430} (2004) 1001.
\bibitem{Kohsaka} Y.~Kohsaka, K.~Iwaya, T.~Hanaguri, M.~Azuma, M.~Takano, and H.~Takagi, Phys. Rev. Lett. {\textbf 93} (2004) 097004.
\bibitem{Michael} M.~Lang and J.~M{\"u}ller: in {\it The Physics of Superconductors}, edited by K.~H.~Bennenmann and J.~B.~Ketterson (Springer-Verlag, Berlin 2003), Vol. II.
\bibitem{Kanoda} K.~Kanoda: Hyperfine Interact. {\textbf 104} (1997) 235.
\bibitem{Kino} H.~Kino and H.~Fukuyama: J. Phys. Soc. Jpn. {\textbf 65} (1996) 2158.
\bibitem{McKenzie1} R.~H.~McKenzie: Science {\textbf 278} (1997) 820.  
\bibitem{Lefebvre} S.~Lefebvre, P.~Wzietek, S.~Brown, C.~Bourbonnais, D.~J\'erome, C.~M\'ezi\`ere, M.~Fourmigu\'e, and P.~Batail: Phys. Rev. Lett. {\textbf 85} (2000) 5420.  
\bibitem{Ito} H.~Ito, T.~Ishiguro, M.~Kubota, and G.~Saito: J. Phys. Soc. Jpn. {\textbf 65} (1996) 2987.
\bibitem{Kawamoto} A.~Kawamoto, H.~Taniguchi, and K.~Kanoda: J. Am. Chem. Soc. {\textbf 120} (1998) 10984.
\bibitem{Sushko} Y.~V.~Sushko, K.~Andres, N.~D.~Kusch, and E.~B.~Yagbuskii: Solid State Commun. {\textbf 87} (1993) 589.
\bibitem{McKenzie2} R.~H.~Mckenzie: Comments Condens. Matter Phys. {\textbf 18} (1998) 309. 
\bibitem{Fournier} D.~Fournier, M.~Poirier, M.~Castonguay, and K.~Truong: Phys. Rev. Lett. {\textbf 90} (2003) 127002.
\bibitem{Limelette} P.~Limelette, P.~Wzietek, S.~Florens, A.~Georges, T.~A.~Costi, C.~Pasquier, D.~J\'erome, C.~M\'ezi\`ere, and P.~Batail: Phys. Rev. Lett. {\textbf 91} (2003) 016401. 
\bibitem{Kagawa} F.~Kagawa, T.~Itou, K.~Miyagawa, and K.~Kanoda: Phys. Rev. B {\textbf 69} (2004) 064511.
\bibitem{Sasaki1} T.~Sasaki, N.~Yoneyama, A.~Matsuyama, and N.~Kobayashi: Phys. Rev. B {\textbf 65} (2002) 060505. 
\bibitem{Sasaki2} T.~Sasaki, I.~Ito, N.~Yoneyama, N.~Kobayashi, N.~Hanasaki, H.~Tajima, T.~Ito, and Y.~Iwasa: Phys. Rev. B {\textbf 69} (2004) 064508.
\bibitem{Mueller} J.~M{\"u}ller, M.~Lang, F.~Steglich, J.~A.~Schlueter, A.~M.~Kini, and T.~Sasaki: Phys. Rev. B {\textbf 65} (2002) 144521.
\bibitem{Akutsu} H.~Akutsu, K.~Saito, M.~Sorai: Phys. Rev. B {\textbf 61} (2000) 4346.
\bibitem{Yoneyama1} N.~Yoneyama, T.~Sasaki, T.~Nishizaki, and N.~Kobayashi: J. Phys. Soc. Jpn. {\textbf 73} (2004) 184.
\bibitem{Yoneyama2} N.~Yoneyama, A.~Higashihara, T.~Sasaki, T.~Nojima, and N.~Kobayashi: J. Phys. Soc. Jpn. {\textbf 73} (2004) 1290.
\bibitem{Taniguchi1} H.~Taniguchi, A.~Kawamoto, and K.~Kanoda: Phys. Rev. B {\textbf 59} (1999) 8424.
\bibitem{Miyagawa} K.~Miyagawa, A.~Kawamoto, and K.~Kanoda: Phys. Rev. Lett. {\textbf 89} (2002) 017003.
\bibitem{Taniguchi2} H.~Taniguchi, K.~Kanoda, and A.~Kawamoto: Phys. Rev. B {\textbf 67} (2003) 014510.
\bibitem{Sasaki3} T.~Sasaki, N.~Yoneyama, N.~Kobayashi, Y.~Ikemoto, and H.~Kimura: Phys. Rev. Lett. {\textbf 92} (2004) 227001.
\bibitem{Yoneyama3} N.~Yoneyama, T.~Sasaki, and N.~Kobayashi: J. Phys. Soc. Jpn. {\textbf 73} (2004) 1434.
\bibitem{Kimura} S.~Kimura, T.~Nanba, T.~Sada, M.~Okuno, M.~Matsunami, K.~Shinoda, H.~Kimura, T.~Moriwaki, M.~Yamagata, Y.~Kondo, Y.~Yoshimatsu, T.~Takahashi, K.~Fukui, T.~Kawamoto, and T.~Ishikawa: Nucl. Inst. Meth. A {\textbf 467-468} (2001) 893.  
\bibitem{Ikemoto} Y.~Ikemoto, T.~Moriwaki, T.~Hirono, S.~Kimura, K.~Shinoda, M.~Matsunami, N.~Nagai, N.~Nanba, K.~Kobayashi, and H.~Kimura: Infrared Phys. Techn. {\textbf 45} (2004) 369.  
\bibitem{Jacobsen} C.~S.~Jacobsen: {\it Semiconductors and semimetals} vol. 27, ed. E. Conwell (Academic Press, Inc. New York, 1988).
\bibitem{Rice} M.~J.~Rice: Solid State Commun. {\textbf 31} (1979) 93.
\bibitem{Eldridge1} J.~E.~Eldridge, K.~Kornelsen, H.~H.~Wang, J.~M.~Williams, A.~V.~S.~Crouch, and D. M. Watkins: Solid State Commun. {\textbf 79} (1991) 583.  

\bibitem{Edep} It is noted that higher frequencies of $\omega_{3}$ observed in the metallic (or insulating) region of the $x =$ 0.6 No.1 sample (Fig. 1(e), $E \parallel c$) in comparison to those of the other samples ($E \parallel a$) are caused by the the polarization dependence of the $\nu_{3}(a_{g})$ mode \cite{Sasaki2}.    

\bibitem{Eldridge2} J.~E.~Eldridge, C.~C.~Homes, J.~M.~Williams, A.~M.~Kini, and H.~H.~Wang: Spectrochimica Acta. {\textbf 51A} (1995) 947.
\bibitem{Cahn} J.~W.~Cahn: J. Chem. Phys. {\bf 42} (1965) 93.
\bibitem{Uehara} M.~Uehara, S.~Mori, C.~H.~Chen, and S.~-W.~Cheong: Nature (London) {\textbf 399} (1999) 560.
\bibitem{Iwaya} K.~Iwaya, Y.~Kohsaka, S.~Satow, T.~Hanaguri, S.~Miyasaka, and H.~Takagi: Phys. Rev. B {\bf 70} (2004) 161103.
\bibitem{Kino2} H.~Kino and H.~Kontani: J. Phys. Soc. Jpn. {\bf 67} (1998) 3691.
\bibitem{Kondo} H.~Kondo and T.~Moriya: J. Phys. Soc. Jpn. {\textbf 67} (1998) 3695 ; J. Phys.: Condens. Matter {\textbf 11} (1999) L363.
\bibitem{Kuroki} K.~Kuroki and H.~Aoki: Phys. Rev. B {\textbf 60} (1999) 3060.
\bibitem{Schmalian} J.~Schmalian: Phys. Rev. Lett. {\textbf 81} (1998) 4232.
\bibitem{Louati} R.~Louati, S. Charfi-Kaddour, A. Ben Ali, R. Bennaceur, and M. H\'eritier: Phys. Rev. B {\textbf 62} (2000) 5957.
\bibitem{Kuroki2} K.~Kuroki, T.~Kimura, R.~Arita, Y.~Tanaka and Y.~Matsuda: Phys. Rev. B {\bf 65} (2002) 100516.
\bibitem{Murakami} S.~Murakami and N. Nagaosa: J. Phys. Soc. Jpn. {\textbf 69} (2000) 2395.
\bibitem{Onoda} S.~Onoda and N. Nagaosa: J. Phys. Soc. Jpn. {\textbf 72} (2003) 2445.
\bibitem{Kawamoto2} A.~Kawamoto, K.~Miyagawa, Y.~Nakazawa, and K.~Kanoda: Phys. Rev. Lett. {\bf 74} (1995) 3455.
\bibitem{Powell} B.~J.~Powell, and R.~H.~McKenzie: Phys. Rev. Lett. {\bf 94} (2005) 047004. 
\bibitem{Gan} J.~Y.~Gan, Y.~Chen, Z.~B.~Su, and F.~C.~Zhang: Phys. Rev. Lett. {\bf 94} (2005) 067005.
\bibitem{Capone} M.~Capone, G.~Saniovanni, C.~Castellani, C.~Di~Castro, and M.~Grilli: Phys. Rev. Lett. {\bf 92} (2004) 106401.
\bibitem{Girlando} A.~Girland, M.~Masino, A.~Brillante, R.~G.~DellaValle, and E.~Venuti: Phys. Rev. B {\bf 66} (2002) 100507.
\bibitem{Varelogiannis} G.~Varelogiannis: Phys. Rev. Lett. {\textbf 88} (2002) 117005. 
\bibitem{Bill} A.~Bill, H.~Morawitz, and V.~Z.~Kresin: Phys. Rev. B {\bf 68} (2003) 144519.
\bibitem{Izawa} K.~Izawa, H.~Yamaguchi, T.~Sasaki and, Y.~Matsuda: Phys. Rev. Lett. {\textbf 88} (2002) 027002.
\bibitem{Arai} T.~Arai, K.~Ichimura, K.~Nomura, S.~Takasaki, J.~Yamada, S.~Nakatsuji, and H.~Anzai: Phys. Rev. B {\textbf 63} (2001) 104518.
\bibitem{Stalcup} T.~F.~Stalcup, J.~S.~Brooks, and R.~C.~Haddon: Phys. Rev. B {\bf 60} (1999) 9309.
\bibitem{YWatanabe} Y.~Watanabe, T.~Shimazu, T.~Sasaki, and N.~Toyota: Synth. Metals {\bf 86} (1997) 1917.
\bibitem{MWatanabe} M.~Watanabe: Ph.D Thesis (Okayama University, 1999) unpublished.  
\bibitem{Kumai} R.~Kumai, Y.~Okimoto, and Y.~Tokura: Science {\textbf 284} (1999) 1645.
\bibitem{Okimoto} Y.~Okimoto, R.~Kumai, E.~Saitoh, M.~Izumi, S.~Horiuchi, and Y.~Tokura: Phys. Rev. B {\textbf 70} (2004) 115104.
\bibitem{Terasaki} K.~Inagaki, I.~Terasaki, H.~Mori, and T.~Mori: J. Phys. Soc. Jpn. {\bf 73} (2004) 3364.
\bibitem{Sasaki4} T.~Sasaki, T.~Fukuda, N.~Yoneyama, and N.~Kobayashi: Phys. Rev. B {\bf 67} (2003) 144521.
\bibitem{Kartsovnik} M.~V.~Kartsovnik: Chem. Rev. {\bf 104} (2004) 5737.
\end{thebibliography}

\end{document}